\documentclass{aa}
\usepackage{graphicx}
\usepackage{rotating}
\usepackage{txfonts}
\newcommand{\mcol}[3]{\multicolumn{#1}{#2}{#3} }
\newcommand{\struut}{\rule[-2ex]{0ex}{5.2ex}}
\newcommand{\struutup}{\rule{0ex}{3.2ex}}
\newcommand{\struutdown}{\rule[-2ex]{0ex}{2ex}}
\newcommand{\cd}{c\,d$^{-1}$}
\begin{document}
   \title{ A photometric study of the light variations of the triple system DG~Leo\thanks{Based 
          on photoelectric observations obtained at San Pedro Mart\'{\i}r, Sierra Nevada 
	  observatories, and CCD observations at Beersel Hills Observatory, as well as on 
	  photoelectric observations with the Belgian {\sc mercator} telescope}
	  }

   \author{ P. Lampens,\inst{1}
            Y. Fr\'emat,\inst{1}
            R. Garrido,\inst{2}
            J. H. Pe\~na,\inst{3}
            L. Parrao,\inst{3}
	    P. Van Cauteren,\inst{4}
	    J. Cuypers,\inst{1}
	    P. De Cat,\inst{5,1}
	    K. Uytterhoeven, \inst{5}
	    T. Arentoft \inst{1,6},
	    and M. Hobart \inst{7}\\
	  }

   \offprints{P. Lampens (Patricia.Lampens@oma.be)}

   \institute{
             Koninklijke Sterrenwacht van Belgi\"e,
             Ringlaan 3, B-1180 Brussel, Belgium
         \and
             Instituto de Astrof\'{\i}sica de Andaluc\'{\i}a, CSIC,
             Camino Bajo de Huetor, 18008 Granada, Spain
         \and
	     Instituto de Astronom\'{\i}a, UNAM
             Apartado Postal 70-264
             M\'exico 04510, D.F.
         \and
	     Beersel Hills Observatory, Laarheidestraat 166, 
	     B-1650 Beersel, Belgium
	 \and
	     Institute of Astronomy, KULeuven
             Celestijnenlaan 200B
             B-3001 Leuven, Belgium
          \and
             Department of Physics and Astronomy, Aarhus University, 
	     DK-8000 Aarhus C., Denmark
	 \and
	     Facultad de F\'{\i}sica, Universidad Veracruzana
             Veracruz, M\'exico \\
 }

   \authorrunning{Lampens et al.}
   \titlerunning{Photometric investigation of the $\delta$ Scuti star DG~Leo}
   \date{Received date/ Accepted date}
   
   \abstract {Multi-site and multi-year differential photometry of the triple star DG~Leo reveals 
   a complex frequency spectrum that can be modelled as the combination of at least three 
   $\delta$ Scuti type frequencies in the range 11.5-13 c/d (with semi-amplitudes of 2-7 mmag) and a 
   superimposed slow variability of larger amplitude. 
   The period of the slow variation fits very well with half the orbital period of the inner spectroscopic 
   binary indicating the presence of ellipsoidal variations caused by the tidally deformed components in a 
   close configuration. 
   These findings, together with the results of a recent spectroscopic analysis (showing that the system 
   consists of a pair of mild Am stars and one A-type component of normal solar composition), infer that 
   DG~Leo is an extremely interesting asteroseismic target. Identification of which component(s) of 
   this multiple system is (or are) pulsating and determination of the excited pulsation modes will 
   both contribute to a much better understanding of the non-trivial link between multiplicity, chemical 
   composition, rotation, and pulsation in the lower part of the classical Cepheid instability strip.

   }
   \maketitle
%
%
  \section{Introduction}

  \subsection{Objectives}
  
  A `bonus' of studying stellar pulsations in binary and multiple systems is that one can exploit 
  the dynamical information to obtain independent constraints on the pulsational models via the 
  determination of the physical properties of the components. Multi-year photometry with a good 
  phase coverage of the beat periods of the oscillations, on the one hand, allows accurate frequency 
  analyses of the combined light. A careful spectroscopic analysis will, on the other hand, 
  provide information on the mass and the luminosity ratios, the effective temperature, the chemical 
  composition, and the surface gravity of each component given that the spectrum is composite. If dynamically 
  determined masses of the system can be obtained with sufficiently good accuracy, these can be directly 
  confronted with the pulsation masses. By comparing the frequency and modal content of the components that 
  originate from the same protostellar environment, one may realistically hope to improve understanding 
  of pulsation physics, since a difference in pulsational behaviour between each can only be attributed 
  to a limited number of (differing) stellar parameters. Binarity/multiplicity may thus lead to an explanation 
  of the so-far unknown amplitude and mode selection mechanism operating in $\delta$ Scuti stars, and to a 
  refined characterization of the pulsations of variable stars in general.
  
  In a review of pulsating $\delta$ Scuti stars in known double or multiple stars, Lampens \& Boffin 
  (\cite{lam01}) discussed several potentially interesting objects, one of which was DG~Leonis which 
  appears to be a triple system whose components are nearly identical. 
  
  \subsection{The target}
  
  DG~Leo (HR~3889, HIP~48218, Kui~44 AB) is an hierarchical triple system which consists of a close 
  double-lined spectroscopic binary (components Aa and Ab) and one distant companion (component B) 
  forming the wider visual binary. It has V = 6.085 mag and (B-V) = 0.25 mag (Mermilliod
  \& Mermilliod \cite{mer98}). The orbital period of the Aa,b system is 4.147 days,
  while the orbital period of the visual pair AB is roughly 200 years (Fekel \& Bopp \cite{fek77}). 
  The visual component B, observed by {\sc hipparcos} with an angular separation of 0.17\arcsec and a 
  magnitude difference of 0.69 mag with respect to Aa,b (ESA 1997), was monitored by means of micrometric 
  observations and speckle interferometry from 1935 till 1997 (Hartkopf et al. \cite{hart2}). The existing 
  astrometric data indicate a highly inclined and possibly eccentric orbit. Both Barnes et al (\cite{bar77}) 
  and Rosvick \& Scarfe (\cite{ros91}) presented evidence for possible shallow eclipses with an inclination 
  of $\approx$ 75$^{\circ}$ for the Aa,b system. 

  Both close components have spectral type A8~IV (Hoffleit \& Jaschek \cite{hof82}), while the composite 
  spectrum was classified as type F0~IIIn (Cowley \cite{cow76}) and type A7~III with
  enhanced Sr (Cowley \& Bidelman \cite{cow79}). All three components of the system are located
  in the $\delta$ Scuti instability strip and are therefore potential candidates for pulsations. 
  Danziger \& Dickens (\cite{dan67}) first reported the short-period variability in the system but were 
  inconclusive about its nature. The claims for short-period oscillations seem to concern mainly one 
  component: component B was classified both as an ultra-short period Cepheid (Eggen \cite{egg79}, with 
  a period $< 0.275$ days) and as a $\delta$ Scuti star (Elliot \cite{ell74}, with a mean periodicity
  of $\rm 0.0818~days$). Possible amplitude and phase changes were also reported (Antonello \& Mantegazza 
  \cite{ant82}; Rosvick \& Scarfe \cite{ros91}). Since these previous studies are based on very limited 
  photometric data sets, little was known about the pulsational characteristics. A first report on the 
  multiperiodicity of the variations was recently presented after performing a preliminary frequency analysis 
  of newly acquired photometric data (Lampens et al. \cite{lam02}). 
  In the following sections, we will present the complete and updated photometric analysis of this interesting 
  asteroseismic target, where metallicity as well as binarity effects may influence pulsational behaviour. 

\section{Observations and data reduction}\label{sec:obs}

\begin{table*}[t]
\label{tab:log}
\setcounter{table}{2}
\begin{center}
\caption[Table 3]{Log of photometric observational campaigns}
\begin{tabular}{cccrrrrc}
\hline
 Dates & Observatory & Observer(s) & Nr &  Comp 1  & Comp 2  & Check & Hrs \struut \\
\hline
Feb.  4 -- 11, 2002 & OAN & JP &  429 & C1 & C2 & no & 40.6 \\
Feb. 11 -- 17, 2002 & OSN & RG &  748 & C1 & C2 & no & 42.0 \\
Feb. 22 -- Mar. 18, 2002 & LPA & PDC \& KU & 90 & C1 & no & no & $^{*}$\\
Jan.  8 -- 14, 2003 & OAN & JP &  269 & C1 & C2 & C3 & 20.0 \\
Jan.  9 -- Mar. 14, 2003 & BHO & PL \& PVC & 1071 & C4 & no & C5 & 58.7 \\
Mar. 17 -- 19, 2004 & OSN & RG &  144 & C1 & C2 & no & 7.0 \\
\hline
\end{tabular}
{\\$^{*}$: non-continuous observations obtained at La Palma observatory (LPA), cf. Sect. 4}
\end{center}
\end{table*}

\subsection{Photoelectric photometry}

During 2002 and 2003, we performed photoelectric photometry campaigns of DG Leo at two prime 
sites located at different longitudes and equipped with identical instrumentation and standard 
Str\"omgren filters in a near simultaneous mode. New uvby data were collected with the 1.5 m 
telescopes at San Pedro Mart\'{\i}r (OAN), M\'exico, between Feb. 4 - 11, 2002 and at Sierra 
Nevada (OSN), Spain, between Feb. 11 - 17, 2002. More uvby data were collected at OAN between 
January 8 -- 14, 2003. 
We used C1 = HD~84497 = HIP~47946 = GSC~1416--210 (G8~III, V=7.42, b-y=0.581) and C2 = HD~84739 = 
HIP~48051 = Cou~284 (F0, V=7.39, b-y=0.290, also a close visual double star) as comparison stars. 
The check star was C3 = HD~86516 = HIP~48977 = GSC~1418--1085 (Am, V=6.74, b-y=0.109). The magnitude 
differences in the sense (C1 - C3) were shown by Rosvick \& Scarfe (\cite{ros91}) to be constant 
at their level of precision. The {\sc hipparcos catalogue} (ESA \cite{esa97}) mentions that C1 
showed no detectable variability throughout the mission, while the comparison stars C2 and
C3 carry the flag 'Duplicity-induced variability'. The standard deviations of the uvby data 
of the OSN 2002 and OAN 2003 campaigns show that C1 and C2 remained constant at the level 
of a few mmag, except in filter u for which the noise level is larger (cf. values between 
brackets in Table~2 introduced later in this Sect.). 
On the other hand, C3 might be variable on a long-term scale, even though it was used for a 
short time as a comparison star by Jerzykiewicz (\cite{jer93}). We found a difference of 
more than 2 $\sigma$ on index m1 and more than 6 $\sigma$ on index c1 when comparing our 
mean indices with the standard indices published by Maitzen et al.~(\cite{mai98}; based on 6 
measurements, see below). 

The observations at OAN were performed according to the sequence 'variable sky C1 variable
C2 ( variable C3 )'. Five successive 10-second integrations per filter were used for the stars, 
while only one 10-second integration per filter was used for the sky. The observations at OSN 
were performed according to the sequence 'variable C1 C2 sky' using 30-second integrations
in each filter for all objects. During the OAN 2003 campaign the check star (C3) was also 
included in the observational sequence. 

All OAN data were reduced according to the same reduction 
method as applied to the OSN data. First, the dead-time correction was applied to all counts. 
Then, subtraction of the (average) sky measurement from the average value of the target 
star was carried out for each cycle. The atmospheric extinction was calculated from the 
comparison star data following the method of Gr{\o}nbech et al. (\cite{gr76}) and the
corresponding corrections were subsequently applied to the magnitude differences. The OAN instrumental 
errors computed for each season using the standard stars are shown in Table~1. 
As can be seen from this table, the OAN instrumental errors were larger for the 2002 than for 
the 2003 season. The reason for the smaller errors in 2003 was the implementation of new phototubes 
for the uvby photometric system in 2002 followed by a debugging phase which lead to more precise 
determination of the dead-time corrections for the 2003 OAN season.

The photometric data discussed here represent magnitude differences in the instrumental system between 
the variable and the comparison star (C1 or C2) interpolated to the observation time of the variable 
star. As the mean differences were not exactly the same for both observatories, we applied small corrective 
shifts for each of the three campaigns (see Table~2). 
Note that the OAN 2003 campaign was held at an epoch closely matching that of the the high-resolution spectroscopic 
run conducted at the Observatoire de Haute-Provence (Fr\'emat et al. \cite{fre2}).
In this way we obtained a homogeneous data set consisting of 1590 epochs and multi-colour differential magnitudes 
of DG Leo. The new material represents about 110 hrs of high-quality photoelectric photometry (Data Set 1).

\subsection{CCD photometry}

In order to increase the amount of data and extend the time basis of the previous data set, 
we also performed complementary observations with a small telescope at Beersel Hills Observatory (BHO), Belgium, 
during the year 2003. Since the target is very bright, we used a 10cm telescope with an SBIG~ST10XMe camera 
to additionally collect differential CCD data, and we adopted C4 = HD~85017 = HIP~48197 = GSC~1417--315 
as the comparison star and C5 = HD~85215 = HIP~48303 = GSC~1417--312 as the check star. 
We also defocused slightly in order to improve the photometric precision. These frames were reduced with 
the aperture photometry procedure of the Mira AP software package\footnote{The software Mira AP is produced 
by Axiom Research Inc., http://www.axres.com/}. 
1071 differential (stacked) V measurements were thus obtained between Jan. 9 and March 14, 2003 representing 
a total of 59 hours of differential CCD photometry in the instrumental system. 
Though the equipment and the filters are not exactly the same and the comparison stars were obviously different, 
we combined all differential V and Str\"omgren y measurements into one large set (Data Set 2 containing 2661 
differential magnitudes in a `single' filter\footnote[2] {The data will be available in electronic form 
via anonymous ftpto {\tt cdsarc.u-strasbg.fr (130.79.128.5)} or via 
{\tt http://cdsweb.u-strasbg.fr/cgi-bin/qcat?J/A+A/xxx/xxx}}), which will serve the purposes of 
a) providing a different time distribution with a distinct alias pattern and b) verifying the 
frequency solution that will be presented in the following section. To perform such data combination, we needed 
to apply small corrective shifts of order 0.01 mag on the CCD differential data computed from the shifts 
between the nightly means of the two comparison stars in the field. The averaged V and Str\"omgren y magnitude 
differences were subsequently subtracted for both data types. No other transformation was applied. 
Table~3 shows the log of the observational runs performed between January 2002 and March 2004.

\begin{table}[t]
\label{tab:errors}
\setcounter{table}{0}
\begin{center}
\caption[Table 1]{OAN errors in the instrumental system and colour indices of the comparison stars 
in the standard system (season 2003)}
\begin{tabular}{ccccc}
\hline
 Season & Filter u & Filter v & Filter b & Filter y \struut \\
\hline
 OAN 2002 & 0.031 & 0.012 &  0.008 &  0.008\struutup \\
 OAN 2003 & 0.024 & 0.005 &  0.006 &  0.006\struutdown \\
\cline{1-5}
 Star &  V ($\sigma_{V}$) & (b-y) ($\sigma_{b-y}$) & m1 ($\sigma_{m1}$) & c1 ($\sigma_{c1}$) \struut \\
\hline
 C1 &  7.394 {\small (20)}&  0.583 {\small (11)}&  0.326 {\small (28)}&  0.346 {\small (45)}\struutup  \\
 C2 &  7.358 {\small (20)}&  0.291 {\small (11)}&  0.144 {\small (28)}&  0.473 {\small (45)} \\
 C3 &  6.729 {\small (20)}&  0.097 {\small (11)}&  0.282 {\small (28)}&  0.563 {\small (45)}\struutdown \\
\hline
\end{tabular}
\end{center}
\end{table}

\begin{table*}[b]
\label{tab:means}
\setcounter{table}{1}
\begin{center}
\caption[Table 2]{Mean differences between variable/comparison 2 and comparison 1 star per season and per filter}
\begin{tabular}{ccccccc}
\hline
 Data type & Campaign & Nr & Filter u & Filter v & Filter b & Filter y \struutup \\
 &&& Mean (St.dev.)& Mean (St.dev.) & Mean (St.dev.) & Mean (St.dev.) \struutdown \\
\hline
 V - C1  & OAN 2002 & 429 &  -2.2251 (0.022) & -2.2572 (0.014) &  -1.6962 (0.012) &  -1.3136 (0.011)\struutup \\
C2 - C1  & OAN 2002 & 420 &  -1.1529 (0.015) & -0.7820 (0.006) &  -0.3107 (0.005) &  -0.0353 (0.004)\\
\cline{2-7}
 V - C1  & OSN 2002 & 748 &  -2.2205 (0.011) & -2.2608 (0.009) &  -1.7454 (0.008) &  -1.3279 (0.007)\\
C2 - C1  & OSN 2002 & 750 &  -1.0932 (0.009) & -0.7351 (0.004) &  -0.3120 (0.003) &  -0.0424 (0.004)\\
\cline{2-7}
 V - C1  & OAN 2003 & 269 &  -2.2216 (0.016) & -2.2835 (0.011) &  -1.7666 (0.010) &  -1.3264 (0.008)\\
C2 - C1  & OAN 2003 & 269 &  -1.1408 (0.011) & -0.7843 (0.002) &  -0.3155 (0.002) &  -0.0340 (0.002)\\
\cline{2-7}
 V - C1  & OSN 2004 & 144 &  -2.2317 (0.010) & -2.2697 (0.008) &  -1.7509 (0.007) &  -1.3336 (0.006)\\
C2 - C1  & OSN 2004 & 144 &  -1.1052 (0.011) & -0.7407 (0.003) &  -0.3160 (0.004) &  -0.0459 (0.003)\struutdown \\
\hline
\end{tabular}
\end{center}
\end{table*}

\section{Analyses}

\subsection{Frequency analysis of Data Set~1}\label{sec:anal1}
  
 We performed the frequency analysis with two different codes that are based on computation 
 of the Fourier Tranform and/or on the technique of least-squares fitting. 
 
 At first, we used {\sc Period98} (Sperl \cite{sprl}), a programme which determines the frequencies one 
 by one by computing the classical Fourier Transform 
 followed by adjustment of the parameters of the periodic functions (zero point, amplitudes, 
 and phases) using least-squares fitting. Next, the computed frequency solution is refined through 
 simultaneous adjustment of all the parameters, including the frequencies in the course
 of a (local) minimization process. Since we are dealing with multiperiodic signals, we used the empirical 
 criterion that a frequency is significant as long as the signal-to-noise ratio of the corresponding 
 semi-amplitude is greater than 4 (Kuschnig et al. \cite{kus97}). 
 
 Secondly, a multi-parameter fitting code that scans a wide range in frequency and searches 
 to minimize the residuals by simultaneously adjusting multiple periodic functions was used 
 (global minimization). In this way accurate, alternative multi-frequency solutions can be found 
 that might be overlooked by the previous code (Schoenaers \& Cuypers \cite{scho}).
 This is especially useful when an alternative solution due to the ambiguity of some aliasing pattern
 is also probable. In this case both computations gave identical results in all filters except
 for filter u which was of lower quality. Table~4 lists the results for Data Set 1 for each filter
 of the Str\"omgren system. 
 Listed are the frequency (in \cd), the semi-amplitude (in mmag), the phase (in 2$\pi$rad), and
 the total fraction of the variance that is removed from the signal (R). In addition we list the 
 signal-to-noise ratio, with respect to the mean noise level in the residual periodogram computed within 
 a box of width 5\cd~centered 
 on the relevant frequency, as well as the standard deviation of the residual signal, $\sigma_{res}$, which
 would correspond if {\sc Period98} were used; and a succession of prewhitenings with the previous frequency 
 would be performed at each step. 

 The dominant frequency F1, the one of largest amplitude found at \@about 0.48 \cd, corresponds to a slow 
 periodic change of the mean intensity linked to the orbital variations of the short-period pair $Aa,b$. It 
 is very close to one half of $P_{orb} = 4.146751~$days, which is the orbital period derived from the published 
 spectroscopic analyses (Fekel \& Bopp \cite{fek77}, Fr\'emat et al. \cite{fre1}). Note that the orbital 
 frequency itself (F0) is not significant, as its signal-to-noise ratio remains far below 4 in all the
 passbands. However, we considered its contribution, since omitting it would affect the other frequency values
 very slightly, particularly in the u passband. Four more frequencies between $11.9~$ and $12.8~$\cd (F2-F5) 
 were detected in each filter. In fact, two of these are located about $0.08~$\cd apart, 
 which caused some caution in the preliminary study where T was only about 12 days (Lampens et al. \cite{lam02}). 
 Their presence is undoubtedly confirmed as the new time base of $767$ days corresponds to a formal resolution 
 of $0.0013$ \cd. 
 Also note that F3 and F4 could be subject to the n.year$^{-1}$ ambiguity (with {\it n} equal to 1 or 3). 
 Obviously, there is excellent agreement between the solutions in all four passbands. We point out the 
 high R-value in filters b and v, but also the poorer match of the same solution in the u passband 
 (R $\sim$ 33\%). From a comparison of the R-values for nearby alias frequency sets, we verified that no 
 other combination was as significant as the one listed for the y passband. As an illustration of 
 this analysis we present 
 Fig.~\ref{fig:anal1}, which is entirely based on the computations performed in parallel with {\sc Period98} and
 derived in successive steps for the data in filter b.
 
\begin{figure}
  \centering
  \includegraphics[angle=0,width=5cm]{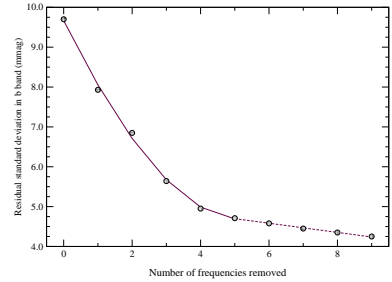}
     \caption{Residual standard deviation as a function of the number of removed sinusoids (Set~1 - filter b)}
    \label{fig:rms}
 \end{figure}

\begin{figure*}
  \centering
  \resizebox{\textwidth}{!}
  {\includegraphics[angle=0]{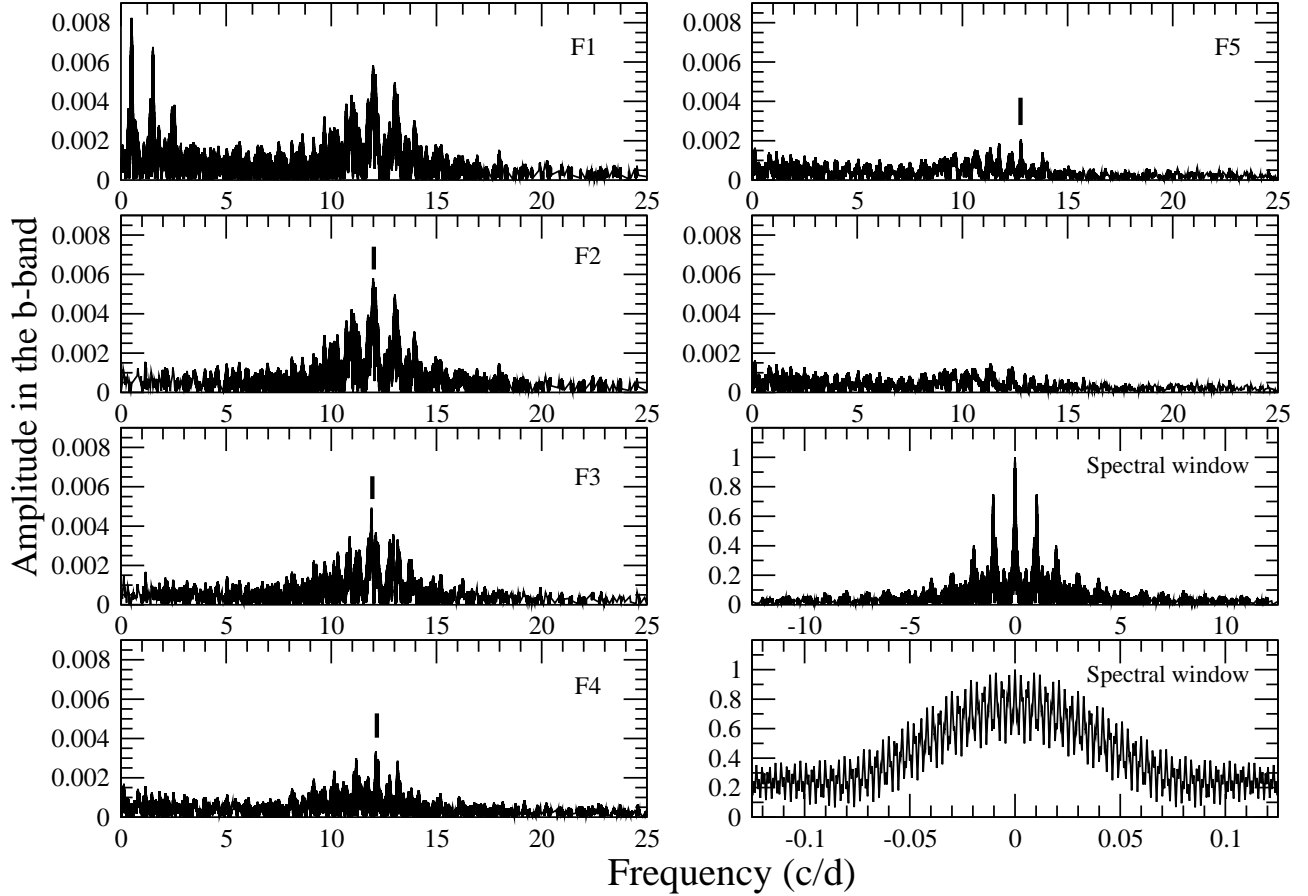}}
    \caption{Successive frequency searches and spectral window of the Fourier analysis (Set~1 - filter b)}
    \label{fig:anal1}
 \end{figure*}

 The expected error in frequency in the case of uncorrelated observations can be computed using  
 the equation \[ \sigma_{F} = \frac{\sqrt{6}}{\pi} . \frac{\sigma_{n}}{A.N^{1/2}.T} \] (Cuypers \cite{cuy87}), 
 where A/$\sigma_{n}$ indicates the signal-to-noise ratio, N, the number of observations and T, the total time base. 
 From Table~4, we adopt 5 mmag as the typical noise level remaining after prewhitening for six frequencies in 
 the filters v, b, and y and derive an error in frequency of 1.-5.10$^{-5}$\cd~on frequencies F1-F5, respectively, 
 given that $T = 767$ days and $N = 1590$ (Data Set~1).
 This also corresponds to the error one would obtain adopting the formula derived by Schwarzenberg-Czerny (\cite{sch91}), 
 provided that some of the simplifications are not considered. As noted by Schwarzenberg-Czerny (\cite{sch91})
 and Montgomery \& O'Donoghue (\cite{mon99}), correlations in the noise may increase this error 
 with the square root of the noise correlation length $\sqrt D$. 
 To estimate D, we analysed the residuals after full prewhitening with six frequencies
 and found a mean sign change at almost every 4 data points.
 In the following we will adopt the conservative error estimates of 2.-10.10$^{-5}$~\cd on frequencies
 F1-F5. In view of the small
 differences obtained between the frequency analyses in the filters y, b, and v, these errors further
 suggest that a unique solution is valid for all the data.
 Apart from these values, which can be considered as upper limits of the errors on the frequencies, we recall the 
 fact that the n.year$^{-1}$ aliasing effect still affects the current analysis and that this is a practical 
 limitation to the true accuracy of the frequencies F3 and F4 in Table~4.
 We decided to stop the frequency search at this level since we noted a significant change of the slope 
 at N~=~4 when plotting the residual standard deviation of the b data as a function of the number of 
 sinusoids removed from the original data (Fig.~\ref{fig:rms}). 

\begin{table}[]
\label{tab:results2}
\setcounter{table}{3}
\begin{center}
\caption[Table 4]{Multi-parameter fits and solutions (Data Set~1)}
\begin{tabular}{cll@{}r@{}rccc}
\hline
 Set && Freq.   & Amp.   & Phase & S/N & $\sigma_{res}$ & R \\
 1  && \cd     & {\tiny mmag} & {\tiny 2$\pi$rad} && {\tiny mmag} & \%\struutdown \\
\hline
   &&{\it Filter y}  &&N=1590  & $\sigma_{init}$= 8.4 mmag&& \struutup\\
 F1&& 0.482282       & 7.7 & 0.41 & 22  & 6.7 &  \\
 (F0&& 0.241141      & 0.8 & 0.36 & 2.2 & - ) &  \\
 F2&& 11.99429       & 5.2 & 0.98 & 13  & 5.8 &  \\
 F3&& 11.91064       & 4.5 & 0.63 & 11  & 5.0 &  \\
 F4&& 12.11436       & 3.2 & 0.00 & 8   & 4.5 &  \\
 F5&& 12.77599       & 1.6 & 0.32 & 4.0 & 4.3 & 74\struutdown \\
   &&{\it  Filter b}  &&N=1590  & $\sigma_{init}$= 9.5 mmag&&\\
 F1&& 0.482276       & 8.4 & 0.41 & 15  & 7.7 &  \\
 (F0&& 0.241138      & 0.4 & 0.31 & 0.8 & - ) &  \\
 F2&& 11.99426       & 6.2 & 0.97 & 13  & 6.5 &  \\
 F3&& 11.91062$^{*}$ & 5.3 & 0.62 & 11  & 5.4 &  \\
 F4&& 12.11439$^{**}$& 3.8 & 0.02 & 8   & 4.7 &  \\
 F5&& 12.77592       & 2.3 & 0.34 & 4.9 & 4.4 & 79\struutdown \\
   &&{\it  Filter v}  &&N=1590  & $\sigma_{init}$= 10.8 mmag&&\\
 F1&& 0.482274       & 8.9 & 0.41 & 14  & 9.0 &  \\
 (F0&& 0.241137      & 0.4 & 0.38 & 0.6 & - ) &  \\
 F2&& 11.99427       & 7.3 & 0.97 & 13  & 7.6 &  \\
 F3&& 11.91062$^{*}$ & 6.4 & 0.62 & 12  & 6.2 &  \\
 F4&& 12.11439$^{**}$& 4.7 & 0.02 & 9   & 5.3 &  \\
 F5&& 12.77591       & 2.6 & 0.34 & 4.8 & 4.9 & 79\struutdown \\
   &&{\it  Filter u}  &&N=1590  & $\sigma_{init}$= 15.5 mmag&&\\
 F1&& 0.482268  & 8.1 & 0.41 & 6   & 14.3 &  \\
 (F0&& 0.241134 & 2.8 & 0.57 & 2.1 & -  ) &  \\
 F2&& 11.99425  & 6.5 & 0.95 & 7   & 13.7 &  \\
 F3&& 11.91055  & 6.3 & 0.59 & 7   & 13.1 &  \\
 F4&& 12.11432  & 3.7 & 0.94 & 4   & 12.8 &  \\
 F5&& 12.77586  & 2.9 & 0.35 & 3.2 & 12.7 & 33\struutdown \\
\hline
\end{tabular}
{\\ $^{*}$ alternatively, the 3.(year)$^{-1}$ alias frequency 11.9197 \cd}
{\\ $^{**}$ alternatively, the 1.(year)$^{-1}$ alias frequency 12.1172 \cd}
\end{center}
\end{table}

\subsection{Frequency analysis of Data Set~2}\label{sec:anal2}

\begin{figure*}[t]
  \centering
  \hspace{2.0cm}
  {\includegraphics[angle=0,width=17.0cm]{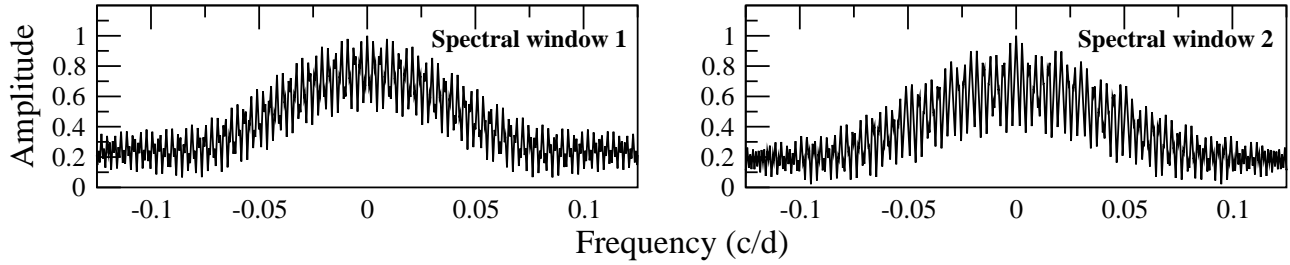}}
    \caption{Spectral windows showing the difference in aliasing between Set 1 and Set 2}
    \label{fig:anal2}
 \end{figure*}

 Data Set 2 spans the same time base as the former data set but consists of 2661 combined measurements 
 in the filters y and V. While the time base is identical, the increased number of measurements should 
 improve the accuracy of the analysis. However, because the site for the CCD data is located 
 near an industrialized area and a smaller telescope was used, the accuracy of a single measurement 
 is lower. We therefore assigned a relative weight of about 5$^{-1}$ for each CCD measurement with 
 respect to a photoelectric one. The inverse ratio of the variances of the differential magnitudes between 
 the comparison stars C4 and C5 (BHO) and C1 and C2 (OSN, OAN) was used to compute this factor. The main 
 asset of these data is that the spectral window is different, largely due to the fact that the CCD observations 
 spanned several months in the 2003 season (cf. Fig.~\ref{fig:anal2}). 
 
 We again performed the same analysis twice: first with {\sc Period98}, then with the method of multi-parameter 
 fitting. We list the best set of parameters when adopting a six-frequency model with the latter method in Table~5.
 From a comparison of the R-values for close sets of aliased frequencies, we also verified that no other 
 combination was as significant as this one. In particular, the frequency at 12.1145 \cd (F4) is slightly more 
 convincing than its year$^{-1}$ alias at 12.1172 \cd. Note that the 3.year$^{-1}$ ambiguity no 
 longer affects the solution. 
 In theory, the formal errors in frequency would be improved by $\sqrt{N_{1}/N_{2}}$ (with N$_{1}$ 
 and N$_{2}$ the number of data in the first and last set respectively). However, considering the 
 use of a relative weight of 5$^{-1}$ for the V data, we will adopt the same conservative errors as before 
 (i.e. errors of 2.-10.10$^{-5}$~c/d respectively on frequencies F1-F5). 
 From here on, we will adopt the frequency solution of the v data (Data Set~1) as the final solution since 
 it is based on homogeneous data of very high quality and since it is furthermore fully consistent with the 
 solution derived for Data Set~2 within the quoted error bars. 
  
\begin{table}[]
\label{tab:results4}
\setcounter{table}{4}
\begin{center}
\caption[Table 5]{Multi-parameter fit and solution (weighted Data Set~2)}
\begin{tabular}{cll@{}r@{}rccc}
\hline
Set && Freq.   & Amp.   & Phase & S/N & $\sigma_{res}$ & R \\
 2  && \cd     & {\tiny mmag} & {\tiny 2$\pi$rad} && {\tiny mmag} &\%\struutdown \\
\hline
  && {\it Filter y+V} && N=2661 & $\sigma_{init}$= 9.1 &&\struutup\\
  F1&& 0.482278      & 7.6  & 0.42 & 13  & 7.5 & \\
 (F0&& 0.241139      & 1.1  & 0.23 & 2.7 & - ) & \\
  F2&& 11.99425      & 5.0  & 0.98 & 13  & 6.7 & \\
  F3&& 11.91064      & 4.0  & 0.63 & 10  & 6.1 & \\
  F4&& 12.11446$^{*}$& 2.9  & 0.00 & 8   & 5.8 & \\
  F5&& 12.77588      & 1.6  & 0.31 & 4.2 & 5.6 & 62\struutdown \\   
\hline
\end{tabular}
{\\ $^{*}$ alternatively, the 1.(year)$^{-1}$ alias frequency 12.1172 \cd}
\end{center}
\end{table}

\subsection{Interpretation of the light variability}\label{sec:lite}
 
 As mentioned, the frequency of highest amplitude has a value close to 2.f$_{orb}$ and corresponds 
 to a slow periodic variability of the daily mean intensity due to the geometrical variations of 
 the inner binary $Aa,b$. With an orbital period as short as 4~days, both components of $Aa,b$ 
 must have their shapes distorted by tidal interaction. We adopted the spectroscopic orbital period 
 (P$_{orb}=4.146751$~days) and used the following model of the light curve of an ellipsoidal binary:
 \begin{equation}
 L_{Aa,b}(\phi) = \alpha + \beta * cos(\phi) + \gamma * cos(2\phi)
 \end{equation}
 with $\phi$ the orbital phase and $\alpha$ (the mean light level), $\beta$, $\gamma$ the coefficients 
 of the fit (Beech \cite{bee89}). The term corresponding to $2*f_{orb}$ (F1) can be associated to the 
 ellipticity effects: the non-spherical cross-sections vary as the system rotates around the centre 
 of mass and maximum luminosity occurs twice per revolution.
 The term corresponding to $f_{orb}$ (F0) is related to the irradiation effect which occurs once per 
 revolution: the inner part of the cooler component is irradiated by the hotter one and re-emits the 
 excess radiation isotropically. There is no net observable effect in this case, meaning that both 
 components
 must have similar effective temperatures (Wilson \cite{wil94}). The question that remains is whether 
 eclipses also occur or not. When fitting the OSN-OAN 2002 data with a single sinusoidal term as a 
 function of half the orbital period, we note the almost constant differences in the nightly means 
 of the OAN run which can be assigned either to shallow eclipses and/or to ellipsoidal variations 
 (Fig.~4), whereas the slow change 
 of the nightly mean levels in the OSN run is evidence for ellipticity effects (Fig.~5).
 This fit also shows that no (shallow) eclipses are needed to explain the observed light variations,
 in contrast to a previous estimation (Rosvick \& Scarfe \cite{ros91}). Thus, the dominant photometric 
 variation can be explained as caused solely by the tidally deformed shapes of both components. 

   \begin{figure}
   \label{figell1}
   \centering
   \includegraphics[angle=-90,width=7.5cm]{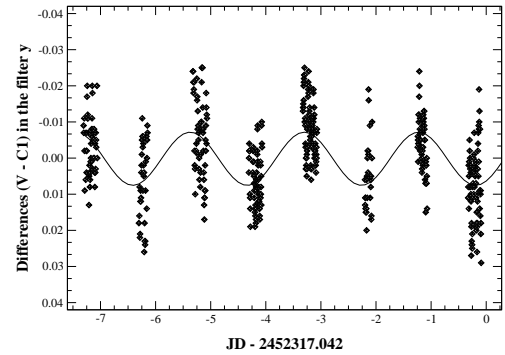}
      \caption{Individual OAN light curves showing the ellipsoidal variation
               of DG~Leo (filter y)}
   \end{figure}
%
   \begin{figure}
   \label{figell2}
   \centering
   \includegraphics[angle=-90,width=7.5cm]{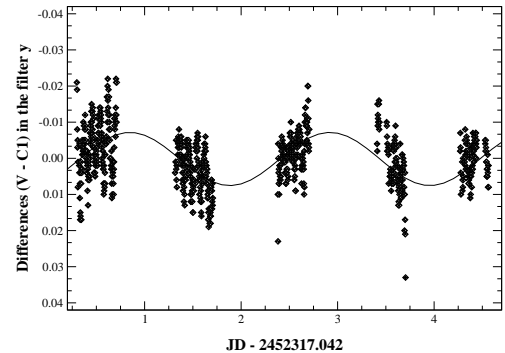}
       \caption{Individual OSN light curves showing the ellipsoidal variation
               of DG~Leo (filter y). Note the different orbital phase coverage.}
   \end{figure}
%

 The other frequencies detected in the light variations may be associated to pulsations of the type 
 $\delta$ Scuti in one (possibly more) component(s) of DG~Leo. Adopting a unique frequency solution 
 (the final one derived in Sect.~3.2) for all passbands, we recomputed the semi-amplitudes and the 
 phases with their standard errors through least squares fitting (Table~6). Note the tiny colour effect 
 in frequencies F1 and F0 as the amplitude in filter u for F0 = f$_{orb}$ becomes larger. The amplitude 
 ratios and phase shifts between the different colours displayed in this table contain information that
 will be helpful for identifying the modes of the excited pulsations (Garrido \cite{gar00}). Note, however, 
 that the true semi-amplitudes will be larger than observed since three almost equally luminous stars were 
 observed jointly. Given that $A_{tr}$ is the true semi-amplitude, $A_{obs}$ the observed one, and f the 
 additional light source with respect to the maximum light of the variable component B, we derive: 
 \begin{equation}
 A_{tr} - A_{obs} = 2.5 \, * \, {\rm log} \frac {1+f.10^{0.4 A_{tr}}}{1+f}
 \end{equation}
 Application of this equation with f = 2 gives the corrected semi-amplitudes 0.021 and 0.006 mag 
 corresponding to the extreme values of 0.007 and 0.002 mag for the observed amplitudes of the pulsation, 
 i.e. a factor of 3 larger. On the other hand, the semi-amplitudes related to the ellipsoidal variations 
 should be corrected by an additional 50\% (applying the same equation with f = 0.5 since the additional 
 light source is component B with respect to components Aa,b).

\begin{table}[]
\label{tab:ampha}
\setcounter{table}{5}
\begin{center}
\caption[Table 6]{Comparison of amplitudes and phases in various passbands }
\begin{tabular}{c@{}@{}c@{}@{}c@{}@{}c@{}@{}c@{}@{}c@{}@{}c@{}@{}c@{}@{}c}
\mcol{1}{c}{Set} & \mcol{2}{c}{Filter u} & \mcol{2}{c}{Filter v} & \mcol{2}{c}{Filter b} 
& \mcol{2}{c}{Filter y \struutup} \\ 
\mcol{1}{c}{1} & \mcol{1}{c}{A$_{1}$} & \mcol{1}{c}{$\Phi_{1}$} & \mcol{1}{c}{A$_{2}$} & \mcol{1}{c}{$\Phi_{2}$} 
& \mcol{1}{c}{A$_{3}$} & \mcol{1}{c}{$\Phi_{3}$} & \mcol{1}{c}{A$_{4}$} & \mcol{1}{c}{$\Phi_{4}$} \\ 
\mcol{1}{c}{} & \mcol{1}{c}{{\tiny mmag}} & \mcol{1}{c}{{\tiny 2$\pi$rad}} & \mcol{1}{c}{{\tiny mmag}} 
& \mcol{1}{c}{{\tiny 2$\pi$rad}} & \mcol{1}{c}{{\tiny mmag}} & \mcol{1}{c}{{\tiny 2$\pi$rad}} & \mcol{1}{c}{{\tiny mmag}} & \mcol{1}{c}{{\tiny 2$\pi$rad} \struutdown} \\ 
\hline
  F1& 8.1  & 0.34 & 8.9  & 0.34 & 8.4  & 0.34 & 7.6  & 0.34 \struutup \\
    & {\tiny $\pm$ 0.5}  & {\tiny 0.06} & {\tiny 0.2}  & {\tiny 0.02} & {\tiny 0.2}  & {\tiny 0.02} & {\tiny 0.2}  & {\tiny 0.02} \\
 (F0& 2.8  & 0.43 & 0.4  & 0.62 & 0.4  & 0.69 & 0.8  & 0.64)\\
    & {\tiny $\pm$ 0.5}  & {\tiny 0.16} & {\tiny 0.2}  & {\tiny 0.43} & {\tiny 0.2}  & {\tiny 0.40} & {\tiny 0.2}  & {\tiny 0.20} \\
  F2& 6.3  & 0.45 & 7.3  & 0.43 & 6.2  & 0.44 & 5.3  & 0.43 \\
    & {\tiny $\pm$ 0.5}  & {\tiny 0.07} & {\tiny 0.2}  & {\tiny 0.02} & {\tiny 0.2}  & {\tiny 0.03} & {\tiny 0.2}  & {\tiny 0.03} \\
  F3& 6.2  & 0.20 & 6.4  & 0.17 & 5.3  & 0.17 & 4.4  & 0.16 \\
    & {\tiny $\pm$ 0.5}  & {\tiny 0.07} & {\tiny 0.2}  & {\tiny 0.03} & {\tiny 0.2}  & {\tiny 0.03} & {\tiny 0.2}  & {\tiny 0.03} \\
  F4& 3.7  & 0.33 & 4.7  & 0.25 & 3.8  & 0.26 & 3.1  & 0.28 \\
    & {\tiny $\pm$ 0.5}  & {\tiny 0.12} & {\tiny 0.2}  & {\tiny 0.04} & {\tiny 0.2}  & {\tiny 0.04} & {\tiny 0.2}  & {\tiny 0.05} \\
  F5& 2.9  & 0.72 & 2.6  & 0.74 & 2.3  & 0.75 & 1.6  & 0.77 \\
    & {\tiny $\pm$ 0.5}  & {\tiny 0.16} & {\tiny 0.2}  & {\tiny 0.07} & {\tiny 0.2}  & {\tiny 0.07} & {\tiny 0.2}  & {\tiny 0.10} \struutdown \\
\hline
\end{tabular}
\end{center}
\end{table}

\section{Determination of global atmospheric properties}\label{sec:atmos}

 We next computed the OAN mean magnitudes and colours of the triple star in the Str\"omgren standard 
 system (for the 2002 season) in order to derive the global atmospheric parameters of the system  
 through application of the calibration procedure (Lester, Gray \& Kurucz \cite{LG86}). The absolute 
 magnitude, effective temperature, surface gravity, and metallicity of DG Leo Aab,B and their standard 
 errors are presented in Table~7. 

\begin{table*}[h]
\label{tab:uvbygen}
\setcounter{table}{6}
\begin{center}
\caption[Table 7]{Mean standard colours and global atmospheric parameters of DG~Leo in two photometric systems}
\begin{tabular}{lcccccccccccc}
\hline
a) Str\"omgren&  V           &   b-y        &    m1         &    c1        & (b-y)$_{0}$  & m$_{0}$      & c$_{0}$      & $\beta$      & M$_{V}$ & T$_{eff}$ & log g & [M/H] \struutup \\ 
              & {\small mag} & {\small mag} & {\small mag } & {\small mag} & {\small mag} & {\small mag} & {\small mag} & {\small mag} & mag     & K         & dex   & dex \struutdown \\
\cline{2-13}
              & 6.0788 & 0.1462 & 0.2210 & 0.9328 & 0.140 & 0.223 & 0.932 & 2.788 & 1.20 & 7540 & 3.59 & 0.32 \\
              & {\small (0.012)}&{\small (0.003)}&{\small (0.005)}&{\small (0.008)} &&&&& {\small (0.30)}&{\small (100)}&{\small (0.12)}&{\small (0.12)} \\
\hline
b) Geneva     & V  & [U-B] & [V-B] & B1-B & B2-B & V1-B & G-B && M$_{V}$ & T$_{eff}$ & log g & [M/H] \struutup \\
              & {\small mag} & {\small mag} & {\small mag } & {\small mag} & {\small mag} & {\small mag} & {\small mag} && {\small mag} & K & dex & dex \struutdown \\
\cline{2-13}
              & 6.0817 & 1.6083 & 0.6467 & 0.9625 & 1.4083 & 1.3656 & 1.7780 && 1.08 & 7340 & 3.90 & 0.21 \\
              & {\small (0.012)}&{\small (0.003)}&{\small (0.006)}&{\small (0.002)}&{\small (0.004)}&{\small (0.005)}&{\small (0.006)}&&{\small (0.31)}&{\small (100)}&{\small (0.12)}&{\small (0.12)} \\
\hline
\end{tabular}
\end{center}
\end{table*}

 We also started to collect absolute data in all seven filters of the Geneva photometric system 
 with the {\sc mercator} telescope\footnote[3]{The {\sc mercator} telescope is operated by the 
 Institute of Astronomy, K.U.Leuven, Belgium, at La Palma (Spain).}. 
 We determined the mean colours and standard deviations of the 90 measurements acquired 
 between February and March 2002 (cf. Table~1). Intrinsic colours of B2 to M0 stars in the Geneva 
 system have been estimated by Meylan et al. (1980) and Hauck (1993). The global colours of 
 DG~Leo seem to match an unreddened MK spectral type of A7~III, except for filter U where 
 the system is less luminous (cf. Table~2 in Hauck \cite{hau93}). We next made use of the 
 {\sc calib} code developed by K\"unzli et al.~(\cite{knz97}) which allows the atmospheric 
 parameters from the Geneva colour indices to be derived for B to mid-G stars on or just above 
 the main sequence. With this code it is also possible to treat metallic-line stars. Using a zero 
 colour excess, we obtained the atmospheric properties listed in Table~7.
 From the reddening-free parameter d = 1.283 (measuring the Balmer discontinuity), we next derived 
 $\Delta$d = 0.107 (measuring how much the star is evolved away from the main sequence) and M$_{V}$ 
 = 1.51 $\pm$ 0.30 mag following Hauck (\cite{hau73}). However, North et al.~(\cite{nor97}) re-examined 
 the existing calibrations of the Str\"omgren and Geneva systems in terms of M$_{V}$ for Am stars. 
 Their conclusion was that the previous calibrations were not reliable and that the Geneva photometry 
 underestimates the luminosity of the hottest Am stars. Applying the new relation on the Str\"omgren 
 indices from Table~7 and using V~sin~{\it i} $\approx$ 30 km/s as an appropriate value (Fr\'emat et 
 al.~\cite{fre2}), we derived the somewhat lower absolute magnitude $M_{V} = 1.08 \pm 0.31$ mag.

 We might compare these calibrated mean values with those that can be directly derived on the basis of 
 the Hipparcos parallax (ESA \cite{esa97}). In Sect.~\ref{sec:lite} we showed that both components of 
 the close binary, Aa and Ab, have similar colours since no net reflection effect is observed in the 
 light curve. We therefore assume that the difference in magnitude between these components is very small 
 (neither one is evolved). On the other hand, we know that the (Hp) magnitude difference between components 
 Aab and B is 0.69 mag (ESA \cite{esa97}). Adopting V = 6.085 mag, we then obtain V$_{Aab}$ = 6.546 mag and 
 V$_{B}$ = 7.236 mag. Similarly we also have V$_{Aa}$ $\equiv$ V$_{Ab}$  = 7.299 mag. Converting to absolute 
 visual magnitudes gives M$_{V_{Aa}}$ $\equiv$ M$_{V_{Ab}}$ = 1.31 mag and M$_{V_{B}}$ = 1.25 mag. 
 The relative parallax error of 15\% contributes the most to the error in such a way that none of the above 
 calibrations can be rejected ($\sigma(M_{V})$ $\approx 2.17 \frac{\sigma_{\pi}}{\pi}$ = 0.33 mag).
 
 The listed errors were increased by 50\% with respect to the nominal accuracies to take the fact into 
 account that three stars were observed jointly. Since the three components were not very different, 
 these derived properties will correspond to approximate values of the physical parameters of the components,
 whereas the high-resolution spectroscopic analysis allowed us to perform a detailed study of the differences 
 between each of the components (cf. Table~6 in Fr\'emat et al. \cite{fre2}). A comparison between these 
 component properties and the global values based on photometric calibrations shows that the agreement in 
 effective temperature is reasonably good (mean of 7500 K), while the agreement in surface gravity is 
 concordant with the value derived from the Geneva photometry (mean of 3.8). Both [M/H] determinations 
 suggest enhanced metallicity of the combined light, an observational fact already mentioned by Fekel 
 \& Bopp (\cite{fek77}) and Cowley \& Bildelman (\cite{cow79}), but only recently quantified 
 (Fr\'emat et al. \cite{fre2}).

 Acquisition of absolute multi-colour photometry in the Geneva photometric system was pursued during 2004 
 with the aim of obtaining well-sampled colour curves for each of the pulsation frequencies. The reduction 
 of these observations is presently taking place. 

\section{Conclusions and future work}\label{sec:con}
 
 The multi-colour time series of DG~Leo recently collected on two continents at the same epoch were 
 merged together and frequency-analysed. Both time series contain high-quality data in the Str\"omgren 
 instrumental system. Their combination furthermore presents the advantage of largely suppressed day$^{-1}$ 
 aliasing effects which made the frequency-analyses straightforward and lead {\small (a)} to an improved 
 interpretation of (the light curve of) the close binary system and {\small (b)} to the detection of 
 multiperiodic oscillations in the combined flux. Monitoring during a few successive years (2002-2004) 
 procured reasonably good accuracy and also permitted verification of the stability of the new frequency 
 solution. 
 
 Up to four $\delta$ Scuti frequencies in the range 11.5-13 \cd\, (with semi-amplitudes of 2-7 mmag in the 
 v passband), as well as a slow variation (with a semi-amplitude of 9 mmag in the v passband) were identified 
 with high confidence. The frequency solution as derived from the Str\"omgren v data appeared to fit the data 
 very well in all other passbands, as well as the more extended data set which is, however, less homogeneous. 
 This unique frequency solution was shown to remain stable over a period of a few years. We further concluded 
 that the period of the slow variation fits very well with half the spectroscopic orbital period of the close 
 binary and that its presence can be explained by continuous ellipsoidal variations supplemented by a tiny 
 colour effect. 
 Despite these efforts, the full frequency content of the pulsations in DG~Leo has not yet been identified 
 as still more peaks are present in the same range of the periodogram of the residuals. A far more extended 
 time series would be needed to determine them with good reliability.
 
 As the triple system cannot be resolved 
 photometrically, this analysis is valid for the combined light of all three components. Previous studies
 indicate that the components have similar characteristics. Therefore, the true semi-amplitudes of the 
 pulsations are diluted by a factor of about 3. The semi-amplitudes and the phases were determined in each 
 of the four Str\"omgren colour indices and will - after finalising the reduction of the {\sc Mercator} data 
 - also be known in the Geneva photometric system. These will be important diagnostics when identifying 
 the degree $\ell$ of the excited pulsation modes. Of course, such analysis depends on correct identification 
 of the component(s) in which the pulsations originate. Only a detailed spectroscopic analysis can provide 
 reliable and accurate information at the component level. Fr\'emat et al. (\cite{fre2}) thus revealed that 
 the close binary consists of two mild Am stars and that line-profile variations on the photometric time-scales 
 of the pulsations were detected in the time-series spectra of the distant companion. The in-depth analysis 
 of DG Leo - a triple system with three almost identical components - should put strict empirical constraints 
 on the development of this type of pulsations in stars of the same mass and age, but of different chemical 
 composition in the outer layers.
 
 Our intention is to postpone the modal interpretation until the frequency analysis of the line profile 
 variations detected in the high-resolution spectra can be presented as well. Reliable conclusions on this 
 issue should indeed come from evidence based on both photometric and spectroscopic diagnostics (both types 
 of data were acquired quasi-simultaneously for a better understanding of this triple system). Only then 
 will we be able to consider the exact origin of the pulsations and to establish the link between 
 multiplicity, chemical composition, and pulsation in this extremely interesting $\delta$ Scuti star.
 A modelling of the pulsations should finally be attempted, in the same way as the preliminary one 
 by Grigahc\`ene et al.~(\cite{gri05}).

\begin{acknowledgements}

 P.~Lampens, Y.~Fr\'emat, and T.~Arentoft gratefully acknowledge financial support from the Belgian 
 Science Policy (through project MO/33/007). 
 R.~Garrido aknowledges financial support from the project ESP2001-4528-EP. L.~Parrao and J.H.~Pe\~na 
 thank the OAN staff for assistance during the observations, Dr.~J.P. Sareyan for measuring the
 OAN dead-time corrections, and PAPIIT $IN110102$ for the funds provided. 
 Some support from the Fund for Scientific Research - Flanders (Belgium) through project G.0178.02 is also 
 acknowledged. Part of these data were acquired with equipment purchased thanks to a research fund 
 financed by the Belgian National Lottery (1999). Extensive use was made of the ADS and Simbad (operated
 by the {\it Centre de Donn\'ees Astronomiques}, Strasbourg, France) data bases. 
 
\end{acknowledgements}

\end{document}